\documentclass[aps,twocolumn,noshowkeys,superscriptaddress,noshowpacs]{revtex4}

\usepackage{dcolumn}
\usepackage{amsmath}
\usepackage{bm}
\usepackage{graphicx}
\usepackage{graphics}

\begin{document}

\title{Valence state and lattice incorporation of Ni in Zn/Co-based magnetic oxides}

\author{V. Ney}
\affiliation{Institut f{\"u}r Halbleiter- und Fest\-k{\"o}r\-per\-phy\-sik, Johannes Kepler Universit{\"a}t, Altenberger Str. 69, 4040 Linz, Austria}
\author{B. Henne}
\affiliation{Institut f{\"u}r Halbleiter- und Fest\-k{\"o}r\-per\-phy\-sik, Johannes Kepler Universit{\"a}t, Altenberger Str. 69, 4040 Linz, Austria}
\author{M. de Souza}
\altaffiliation{Present address: IGCE, Unesp - Univ Estadual Paulista, Departamento de F\'{i}sica, 13506-900, Rio Claro, SP, Brazil}
\affiliation{Institut f{\"u}r Halbleiter- und Fest\-k{\"o}r\-per\-phy\-sik, Johannes Kepler Universit{\"a}t, Altenberger Str. 69, 4040 Linz, Austria}
\author{W. Jantsch}
\affiliation{Institut f{\"u}r Halbleiter- und Fest\-k{\"o}r\-per\-phy\-sik, Johannes Kepler Universit{\"a}t, Altenberger Str. 69, 4040 Linz, Austria}
\author {K. M. Johansen}
\affiliation{Department of Physics, Centre for Materials Science and Nanotechnology, University of Oslo, N-0316 Oslo, Norway}
\author{F. Wilhelm}
\affiliation{European Synchrotron Radiation Facility (ESRF), CS 40220, 38043 Grenoble Cedex, France}
\author{A. Rogalev}
\affiliation{European Synchrotron Radiation Facility (ESRF), CS 40220, 38043 Grenoble Cedex, France}
\author{A. Ney}  
\email{andreas.ney@jku.at}
\affiliation{Institut f{\"u}r Halbleiter- und Fest\-k{\"o}r\-per\-phy\-sik, Johannes Kepler Universit{\"a}t, Altenberger Str. 69, 4040 Linz, Austria}

\begin{abstract}

Ni incorporation has been studied in a comprehensive range of Zn/Co-based magnetic oxides to elucidate it valence state and lattice incorporation. The resulting structural and magnetic properties are studied in detail. To the one end Ni in incorporated by in-diffusion as well as reactive magnetron co-sputtering in wurtzite ZnO where only the Ni-diffused ZnO exhibits significant conductivity. This is complemented by Ni and Co codoping of ZnO leading. To the other end, the ZnCo$_2$O$_4$ spinel is co-doped with varying amounts of Ni. In the wurtzite oxides Ni is exclusively found on tetrahedral lattice sites in its formal 2+ oxidation state as deep donor. It behaves as an anisotropic paramagnet and a limited solubility of Ni about 10\% is found. Due to its smaller magnetic moment it can induce partial uncompensation of the Co magnetic moments due to antiferromagnetic coupling. In the spinel Ni is found to be incorporated in its formal 3+ oxidation state on octahedral sites and couples antiferromagnetically to the Co moments leading again to magnetic uncompensation of the otherwise antiferromagnetic ZnCo$_2$O$_4$ spinel and to ferrimagnetism at higher Ni concentrations. Increasing Ni even further leads to phase separation of cubic NiO resulting in an exchange-biased composite magnetic oxide.  

\end{abstract}

\maketitle

\section{Introduction}

Zn/Co-based magnetic oxides in its various forms have been investigated since decades. To one end, ZnO, which grows in the hexagonal wurtzite structure, usually contains a range of 3$d$ impurities. One of them is Co$^{2+}$, which can be easily identified via its characteristic hyperfine octet in electron spin resonance (ESR), e.\,g., \cite{JGH07}. The magnetic properties of Co impurities in ZnO can be described by a well-established effective spin Hamiltonian introduced for the evaluation of ferromagnetic resonance \cite{EdW61} and optical investigations \cite{Koi77}. The uniaxial single ion anisotropy $DS^2_z$ was experimentally found to be $D/k_B=4$\,K \cite{EdW61,Koi77} leading to a magnetically soft $a$-plane. This was later-on also found in Zn$_{1-x}$Co$_x$O (Co:ZnO) in the range of $x \sim 0.01 - 0.1$ \cite{SHK06,NKO10}, a so-called dilute magnetic semiconductor (DMS). Co:ZnO was under discussion for ferromagnetic order at room temperature \cite{Cha10,Oga10}, however it appears to be meanwhile settled, that phase-pure Co:ZnO remains an anisotropic paramagnet \cite{SHK06,NKO10,NOY08,NOK10} below the coalescence limit of the cationic sublattice of $x\sim 0.2$ \cite{LMZ00}, where next-cation neighbor antiferromagnetic interactions lead to a partial compensation of the Co magnetic moments as well as an effectively reduced single ion anisotropy \cite{NNW12,NHL16}. Co:ZnO above the coalescence-limit thus behaves like an uncompensated antiferromagnet, which exhibits a vertical exchange-bias effect \cite{HNS16}. Throughout the concentration series from the impurity limit up to $x=0.6$, Co is present in its 2+ formal oxidation state and is substitutionally incorporated on Zn lattice sites (Co$^{2+}_{Zn}$). This configuration has a well-established spectroscopic signature in x-ray linear dichroism (XLD) and x-ray absorption near edge spectroscopy (XANES) \cite{NKO10,NOK10,NHL16}.

To the other end there exists ZnCo$_2$O$_4$ which grows in the cubic spinel structure of the generic AB$_2$O$_4$ form. In case of a normal spinel structure the A cation (Zn) is located on the tetrahedrally coordinated sites (Th) and the B cation (Co) on octahedrally coordinated lattice sites (Oh). ZnCo$_2$O$_4$ has been discussed for a range of potential applications, e.\,g., as anode material for lithium batteries~\cite{SSR07}, as $p$-type gate in junction field-effect transistors~\cite{SWG12}, or more recently as efficient electrocatalyst for the oxygen evolution reaction \cite{DAC21}. In general, ZnCo$_2$O$_4$ is a known $p$-type conducting oxide material in which Co is incorporated in its 3+ formal oxidation state at octahedral sites (Co$^{3+}_{Oh}$), while the Zn is located at tetrahedral sites as Zn$^{2+}$ (Zn$^{2+}_{Th}$) \cite{Cos56}. Zn$_{Oh}$ anti-site defects are suspected to act as active acceptors and to produce the holes responsible for the observed $p$-type conductivity \cite{PPZ11}. Also deviations from the ideal stoichiometric composition are considered as source for $p$-type conductivity \cite{DAC21}. It has been shown, that it is possible to grow both wurtzite Zn$_{0.4}$Co$_{0.6}$O (60\% Co:ZnO) as well as spinel ZnCo$_2$O$_4$ and to adjust the resulting phase only by the preparation conditions; however, the magnetic properties of ZnCo$_2$O$_4$ remained mostly paramagnetic \cite{HNO15}. On the other hand, the Co$_3$O$_4$ spinel is known to be an antiferromagnet with a N\'eel-temperature of 40\,K \cite{Rot64}. It may thus be expected that the ZnCo$_2$O$_4$ spinel can also be an uncompensated antiferromagnet, where the degree of compensation depends on the Zn$_{Oh}$ anti-site defect concentration and/or the deviation from the ideal stoichiometry. 

Theoretical predictions of the existence of ferromagnetism of Ni-doped ZnO based DMS (Ni:ZnO) could be found alongside with those for Co:ZnO \cite{SKY00}. First experimental results for ferromagnetism in Ni:ZnO nanocrystals were made shortly after \cite{RaG03} but until today most of these claims are restricted to nanocrystals of ZnO, see, e.\,g., \cite{PXC06,DBA15,SKS19}. In most cases the structural as well as the magnetic characterization is limited by the nanostructured nature of the samples. The changes in optical properties upon Ni-doping have been studied in more detail in \cite{LLC14,PSG15,GSM16}. There are reports on very low Ni concentrations below 1\% \cite{GSM16}, reports on higher Ni concentrations of up to 10\% \cite{LLC14,GSM16}---or even 30\% \cite{DBA15}---which all still claim phase pureness. However, there are also reports on the onset of a structural transformation of ZnO starting from 5\% of Ni doping where the Ni starts to be not fully incorporated on substitutional sites anymore \cite{GoS13}. It is noteworthy, that Ni:ZnO is not only of interest because of its optical or magnetic properties but also because of its antibacterial effects \cite{VMP16} and, like other transition metal dopants, for its photocatalytic activity \cite{KPP18,MGS19}. 

Here we present a detailed experimental study on Ni-doping of wurtzite ZnO as well as of Ni-codoping of the Zn/Co-based magnetic oxides from wurtzite to spinel. The actual valence state as well as the incorporation of the Ni into the host lattice shall be investigated by a range of experimental techniques and will be corroborated by simulations of the respective XANES and XLD spectra as done before for Co:ZnO \cite{NOY08} and Co/Cu-codoped ZnO \cite{NVW19}. It will turn out that mostly Ni$^{2+}$ on Zn lattice sites is formed in wurtzite ZnO which results in a single ion anisotropy with an easy $c$-axis which is opposite to the easy $a$-plane of Co:ZnO. Ni impurities can be photo-ionized while at higher Ni concentrations a pronounced conductivity appears due to secondary mechanisms. The solubility of Ni in ZnO is lower than for Co and is limited to about 10\%. Due to its smaller magnetic moment, Ni codoping of Zn/Co-based oxides can also lead to a partial uncompensation of the effective magnetic moment. Ni-incorporation into the ZnCo$_2$O$_4$ spinel leads to significant changes in the entire lattice occupation and to drastic changes in the magnetic properties. However, under idealized conditions Ni$^{2+}_{Td}$ is formed in the wurtzite lattice while Ni$^{3+}_{Oh}$ dominates in the spinel lattice. In both cases an antiferromagnetic coupling to the adjacent Co ions is favored. Finally, we can provide first indirect evidence for the formation of Zn$_{Oh}$ anti-site defects in the spinel by characteristic spectroscopic signatures in the XANES.    

\section{Experimental details}

Ni-doped ZnO as well as wurtzite and spinel Zn/Co oxide thin films were epitaxially grown on $c$-plane sapphire [Al$_2$O$_3$(0001)] single crystal substrates using reactive magnetron sputtering (RMS). For Ni-doped ZnO and Ni/Co-doped ZnO in the wurtzite structure metallic composite targets (Ni/Zn and Ni/Co/Zn) with a nominal Ni (and Co) content of 10\%, respectively, were sputtered at a low power of 10~W. The base pressure of the preparation system was $2\times10^{-9}$ mbar, while the working pressure during film deposition was $4\times 10^{-3}$ mbar. The composition of the sputter gas is controlled via individual mass flow controllers keeping an Ar:O$_2$-ratio of 10:1 standard cubic centimeters per minute (sccm). The 10\% Ni:ZnO with best structural quality was obtained at a substrate temperature of 300$^{\circ}$C while for 10\% Ni:10\% Co:ZnO it had to be increased to 500$^{\circ}$C. Additional films were grown in a triple cluster magnetron sputtering system with a base pressure of $3\times10^{-8}$ mbar from individual Ni, Co, and Zn metallic targets where the composition of the sample could be controlled via the sputtering power for each magnetron. For comparison, a 10\% Ni:10\% Co:ZnO film was grown with an Ar:O$_2$-ratio of 10:2 sccm at a substrate temperature of 375$^{\circ}$C; a 1\% Ni:ZnO film could be grown under identical preparation conditions. Altering the preparation conditions in the triple cluster system and adjusting the individual sputter-rates it was also possible to grow Zn(Co/Ni)$_2$O$_4$ spinel films. For this the optimum preparation conditions were an Ar:O$_2$-ratio of 10:4.5 sccm and a substrate temperature of 300$^{\circ}$C. The sputtering power was kept at 50~W for Co and 5~W for Zn, while Ni was varied from 5 to 20 to 35~W; in the latter case Zn was reduced slightly to 3~W. Note that for the metallic Zn target even at these low powers the achievable rates are rather large, while for the other targets rather high powers are needed for moderate rates. All sputtering rates of the individual targets were checked via a quartz-crystal micro-balance prior to the deposition. The nominal film thickness throughout this work was 200~nm; x-ray reflectivity (XRR) measurements on selected samples reveal that this corresponds to $150\pm15$~nm thick films. To analyze the global structural properties, x-ray diffraction (XRD) measurements were done with a PANalytical Xpert MRD XL diffractometer to record $\omega - 2 \theta$ and XRR scans. To introduce Ni into ZnO from the gas phase ZnO bulk single crystals (Tokyo Denpa) samples were sealed in quartz-ampoules together with pieces of a Ni-wire of purity 99:995\%, followed by a heat treatment at 1050$^{\circ}$C and 1100$^{\circ}$C for 90~min, respectively. The in-diffusion of Ni extends to about 6 microns into the ZnO as confirmed by secondary ion mass spectroscopy (SIMS). The total Ni dose as in-diffused in both the Zn- and O- terminated surfaces is $7.2 \cdot 10^{15}$/cm$^2$ and $1.3 \cdot 10^{16}$/cm$^2$, referred to as "low" and "high" Ni-content, respectively.    

The x-ray absorption near edge spectra (XANES) were taken at the ID12 beamline of the ESRF in total fluorescence yield in back-scattering geometry \cite{Rog13}. X-ray linear dichroism (XLD) spectra at the Ni, Co and Zn $K$-edges were measured at 300~K as the direct difference of normalized XANES recorded under 10$^{\circ}$ grazing incidence with two orthogonal linear polarizations. A quarter wave plate was used to flip the linear polarization of the synchrotron light from vertical to horizontal, i.\,e., the $E$ vector of the synchrotron light was either parallel or perpendicular to the $c$-axis of the ZnO epitaxial film. The isotropic XANES was derived from the weighted average of the two spectra, i.\,e., ($2 \times$ XANES($E \perp c$) $+$ XANES($E \parallel c$)$)/3$. During one beamtime the quarter wave plate was not available, so that XANES with circular polarized light under grazing and normal incidence had to be recorded instead. This is known to yield identical XLD spectra \cite{NNK14}, however, on the expense of a reduced signal-to-noise ratio and a distorted background which both become apparent for low XLD signals. The x-ray magnetic circular dichroism (XMCD) measurements were taken as the direct difference of XANES spectra recorded with right and left circular polarized light under grazing incidence ($10-15^\circ$). The XMCD spectra were recorded in an external magnetic field of up to 17~T provided by a superconducting magnet. To minimize artifacts, the external field was reversed as well. The CoO reference XANES was taken under grazing incidence using circular polarized light on a powder sample; the XANES was corrected for self-absorption effects. 

$X$-band (9.5~GHz) Electron Paramagnetic Resonance (EPR) spectra were recorded between room temperature and 3~K employing a Bruker spectrometer combined with a liquid Helium flow cryostat. Field modulation of the external field was used for lock-in detection. In addition, the samples could be illuminated by a white-light source (Olympus Highlight 300), as well as with various light emitting diodes from the infrared to ultraviolet regime (photo-ESR). For the integral magnetic characterization $M(H)$ curves were taken at 2~K and 300~K using superconducting quantum interference device (SQUID) magnetometry (Quantum Design MPMS XL-5) applying the magnetic field perpendicular to the $c$-axis of the film (in the plane of the film) as well as parallel to the $c$-axis (out-of-plane). The diamagnetic contribution of the sapphire substrate has been derived from the $M(H)$ curve recorded at 300~K between 2 and 5~T and subtracted from all data. Great care has been taken to avoid the various measurement artifacts for this kind of samples \cite{SSN11,BHH18}. $M(T)$ curves were measured in a field of 10~mT while warming from 2~K to 300~K after the sample was either cooled down from 300~K to 2~K in 5~T (FH) or in zero-field (ZFC). After the ZFC curve another $M(T)$ curve was measured while cooling down in 10~mT (FC). For selected samples $M(H)$ curves were measured at 2~K after cooling the films from room temperature to 2\,K in $+5$~T (plus field cooled - pFC) or $-5$~T (minus field cooled - mFC) external magnetic field as well as ZFC, analogous to previous studies concerning the vertical exchange bias effect in Co:ZnO \cite{HNS16}. 

\section{Results and Discussion}

\subsection{Ni-diffused ZnO single crystals}

\begin{figure}[tb]
\resizebox{1\columnwidth}{!}{\includegraphics{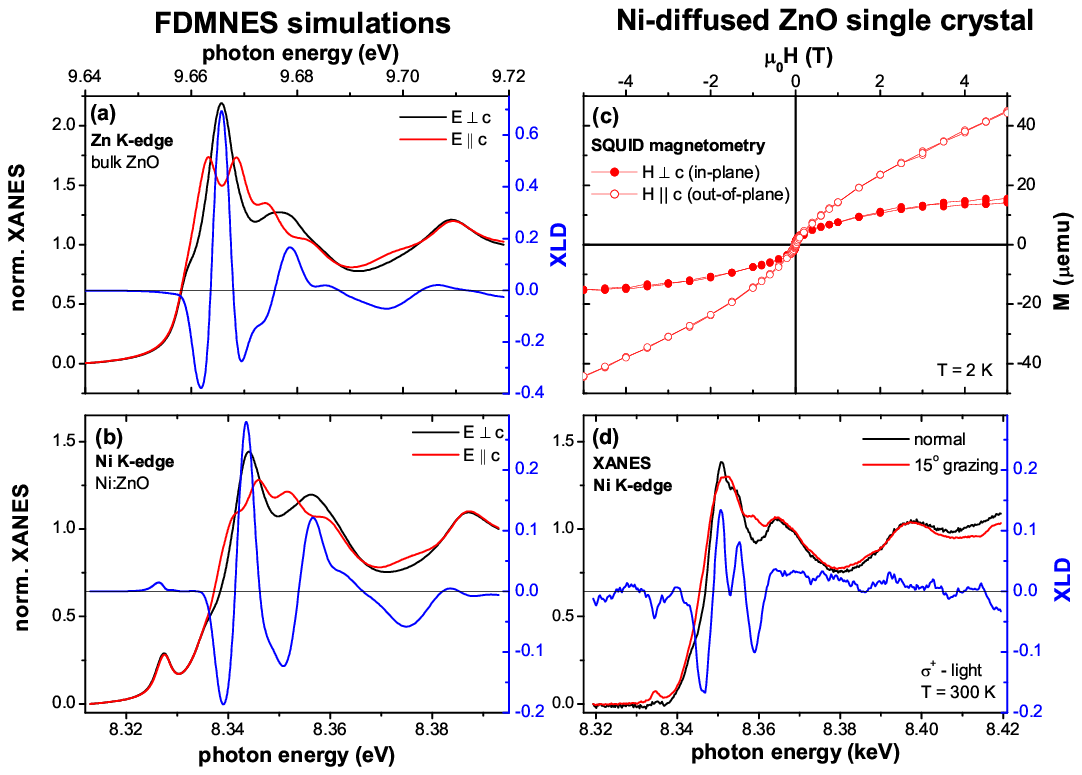}}
\caption{FDMNES simulations of the XANES and XLD at the Zn (a) and Ni (b) $K$-edge for substitutional Ni in wurtzite ZnO. (c) In-plane and out-of-plane $M(H)$ curves of a Ni-diffused ZnO measured at 2~K using SQUID magnetometry. (d) XANES and XLD spectrum measured at the Ni $K$-edge at room-temperature.  \label{fig1}}
\end{figure}

XANES and XLD spectra have been proven to be a suitable tool to study the valence and lattice-incorporation of dopants in wurtzite ZnO. To get an idea about the characteristic spectral signatures of the Ni dopants which are incorporated substitutionally on Zn lattice sites, FDMNES simulations are carried out. For that a (2x2x3) supercell of 1 Ni and 23 Zn atoms as cationic hcp sublattice and 24 O atoms as anionic hcp sublattice was constructed which has the bulk lattice parameter ($a = 3.2459$~\AA\ and $c = 5.2069$~\AA) and a dimensionless $u$-parameter of 0.382 for wurtzite ZnO \cite{KiE89}. The simulated spectra were convoluted with a core hole lifetime (FWHM) of 1.67~eV for Zn \cite{KrO79} and an arctangent-like energy-dependent broadening as done before \cite{NOY08}. The identical broadening parameters were used for the Ni spectra, however for Ni the core hole lifetime was changed to 1.44~eV \cite{KrO79}. The resulting XANES and XLD spectra are shown in Fig.\ \ref{fig1} for the Zn (a) and Ni (b) $K$-edges, respectively. The Zn $K$-edge exhibits the well-known signature of wurtzite ZnO. The Ni $K$-edge XLD has a quite similar appearance to the Zn one, which reflects the fact, that the local crystal field around the absorbing atom, which is anisotropic in case of wurtzite ZnO, is rather similar in both cases and dominated by the tetrahedral arrangement of the surrounding O anions.  In addition the XLD spectrum for substitutionally incorporated Ni compares well with previously reported simulations \cite{SGJ12} and in addition very similar to the previously reported ones for substitutional Co \cite{NOY08} or Cu \cite{NVW19}, underlining the structural origin of the XLD. Note, that the simulations also include the so-called pre-edge feature of the $K$-edge which originates from 3$d$-4$p$ hybridized states and is typically found to be more pronounced in the simulations compared to the actual experiment

In Fig.\ \ref{fig1} (c) the $M(H)$-curves of a Ni-diffused ZnO single crystal is shown, which was recorded at 2~K applying the external magnetic field either in the film plane (open symbols) or out-of-plane (full symbols). The data show a paramagnetic signal, i.\,e., no significant hysteresis is visible within the experimental accuracy, but a pronounced anisotropy of the $M(H)$ curves is visible with a much more pronounced magnetic response out-of-plane, i.\,e., along the $c$-axis of the ZnO lattice. Such a single ion anisotropy is known, e.\,g., for Co:ZnO \cite{SHK06,NKO10}; however, while Co:ZnO exhibits a single ion anisotropy with an easy $a$-plane, for Ni:ZnO an easy $c$-axis behavior is found. The presence of a single ion anisotropy is a first, indirect evidence for a substitutional incorporation of the in-diffused Ni. Figure \ref{fig1} (d) provides more direct experimental evidence for the substitutional incorporation of Ni by means of the experimental XANES and XLD spectra recorded at the Ni $K$-edge. The overall shape and size of the Ni $K$-edge XLD in very comparable to previously reported for Ni ion-implnted samples \cite{SGJ12}. However, the present experiments fall into a beamtime where the quarter wave plate was not available so the XLD had to be derived from XANES recorded using circular polarized light under grazing and normal incidence. Despite the slightly distorted background originating from the altered detection geometry for in-plane and grazing, a clear XLD signature can be seen which approximately follows the characteristic features of the simulations in terms of sign, size and position of the various peaks with the exception of the first pronounced positive peak at the main absorption which is nevertheless consistent with previous experimental findings \cite{SGJ12}. Unlike for Co:ZnO, where the match between simulation and experiment was striking \cite{NOY08}, here the overall agreement is less so that one can conclude that only a significant fraction of the Ni is substitutionally incorporated on Zn lattice sites, presumably with some local distortions, while other Ni species cannot be fully ruled out. 

\begin{figure}[tb]
\resizebox{1\columnwidth}{!}{\includegraphics{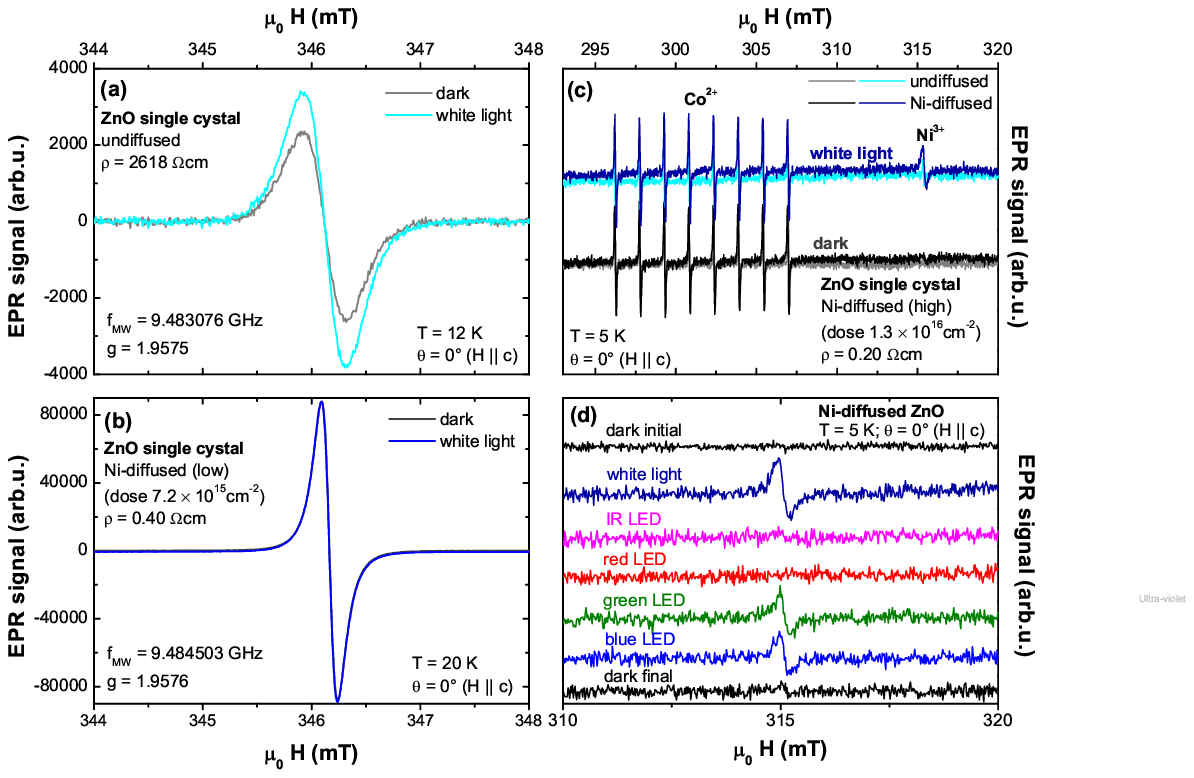}}
\caption{EPR spectrum in the range of the shallow donor of the undiffused (a) and low Ni-diffused (b) ZnO single crystal at low temperature with and without white-light illumination. (c) EPR spectrum in the region of Co$^{2+}$ and Ni$^{3+}$ at 5 K for the undiffused and highly Ni-diffused ZnO sample with and without white-light illumination. (d) Ni$^{3+}$ EPR-line upon illumination using different photon energies. \label{fig2}}
\end{figure}

The Ni-diffused ZnO single crystals also show a significantly altered electronic behavior. While the as received ZnO has a rather high resistivity of 2618~$\Omega$cm, the low and high Ni diffused samples exhibit a much lower resistivity of 0.40~$\Omega$cm and 0.20~$\Omega$cm, respectively. Since one can expect that the substitutional Ni is present in its 2+ formal oxidation state, it should be isoelectric, i\,e., it should not directly act as donor or acceptor in ZnO when substitutionally incorporated in ZnO. Thus, the mechanism behind the increased conductivity upon Ni diffusion is most likely an indirect one. Here we only want to focus on the properties of the Ni itself. For that we take Ref.\ \cite{JGH07} as starting point since a comparable ZnO single crystal from Tokyo Denpa has been investigated using EPR with and without illumination. It is known, that even the undiffused ZnO contains a range of paramagnetic impurities among them are Co$^{2+}$ and Ni$^{2+}$ and/or Ni$^{3+}$. Note, that Ni$^{2+}$ cannot be detected in X-band EPR. It was reported that the Ni$^{3+}$ is present in the ZnO as evidenced by a sharp EPR line around 316~mT for X-band frequencies; this line is close to the characteristic hyperfine-split octet of Co$^{2+}$ impurities and was reported to disappear under illumination with 325~nm light \cite{JGH07}. Illumination also influences a broader EPR line around 346~mT ($g=1.96$) which was associated with a shallow donor in \cite{JGH07}. Figure \ref{fig2} summarizes the EPR investigations of the undiffused ZnO single crystal in comparison with the Ni-diffused ones. Figure \ref{fig2}(a) shows the EPR line of the shallow donor of the undiffused ZnO with and without illumination with white light and the intensity of this line increases slightly with illumination suggesting photo-activated carriers. The low Ni-diffused ZnO in Fig.\ \ref{fig2}(b) exhibits the identical line of the shallow donor, however with a much larger intensity compared to the undiffused ZnO. The smaller linewidth of the doped sample indicates motional narrowing due to the increasing delocalization of bound donor electrons as it was recently also inferred in \cite{SVK20}. Although the actual settings of the EPR spectrometer were not fully identical for the two experiments, the apparent increase of the intensity by around a factor of ten is significant and also corroborated by the increased signal-to-noise ratio for the Ni-diffused sample. In contrast to the undiffused sample, no significant increase of the intensity upon white-light illumination is noticeable. The Ni-diffused sample with the higher dose shows an even more increased shallow donor line (not shown). In that regard the present findings report a significant increase of the shallow donor upon Ni diffusion which is associated with a strong increase in conductivity. It should be noted that the effect of the Ni diffusion on the behavior of the shallow donor can be rather complex. Other unintentional dopants such as, e.\,g., Al and defect complexes involving Zn vacancies are reported to be present in nominally undoped ZnO and they are also sensitive to illumination \cite{SJB14} so that those vacancies may be filled by interstitial Zn which may be generated by the Ni diffusion. However, the identification of the actual mechanism behind the increase of the shallow donor upon Ni diffusion goes beyond the scope of this paper. 

With regard to the Ni itself our findings are in contrast to the ones in \cite{JGH07}. The undiffused ZnO does not exhibit any significant amount of Ni$^{3+}$ in the dark state which can be seen from Fig.\ \ref{fig2}(c), grey line. While Co$^{2+}$ is nicely visible, no Ni$^{3+}$ EPR line can be observed although the spectra were recorded at virtually the same temperature of 5~K (vs. 4.8~K) as in \cite{JGH07}. Only after illumination with white light (cyan line) a tiny Ni$^{3+}$ EPS line is visible. Obviously the unintentional doping of the hydrothermally grown ZnO is not always identical from batch to batch and in the present case at least part of the increase in the shallow donor line can be associated with a photo-ionization of Ni$^{2+}$. Another part of the increase in the shallow donor line can be attributed to the slight reduction of the intensity of the Co$^{2+}$ octet upon white light illumination which is consistent with earlier findings \cite{IGD15}. Upon Ni-diffusion the behavior of the Ni ESR line only changes slightly. In the absence of illumination (black line) there is still no significant amount of Ni$^{3+}$ visible, while after white-light illumination a  Ni$^{3+}$ line appears which is more pronounced in comparison to the undiffused ZnO. The energetic position of the Ni$^{2+}$ defect level can be estimated based on the illumination with selected LEDs shown in Fig.\ \ref{fig2}(d). The appearance and disappearance of the Ni$^{3+}$ line is reversible and cannot be induced by infrared or red light, while with green and blue light the line appears. Therefore, the energetic position of the Ni-level should be of the order of 2.5~eV below the conduction band which experimentally corroborates theoretical predictions of Ni in ZnO being a deep donor \cite{RLZ09}. Note that in agreement with \cite{JGH07} the line disappears upon irradiation with ultraviolet light, i.e. above the band-gap (not shown).

Summarizing this part, the Ni-diffused ZnO samples exhibit significantly increased conductivity which is however not directly induced by the Ni dopants. The Ni itself is (mostly) incorporated substitutionally on Zn lattice sites in its formal 2+ oxidation state, at least down to EPR sensitivities no Ni$^{3+}$ can be detected. Irradiation with white light leads to photoionization of the Ni which is reversible, i.\,e., the Ni$^{3+}$ defect immediately traps any available free carrier in this rather conducting samples once the light is switched off. The Ni dopant exhibits anisotropic paramagnetism with a single ion-anisotropy favoring the $c$-axis which is opposite to the Co$^{2+}$-impurity with an easy $a$-plane. 

\subsection{Wurtzite Ni:ZnO epitaxial films}

\begin{figure}[tb]
\resizebox{1\columnwidth}{!}{\includegraphics{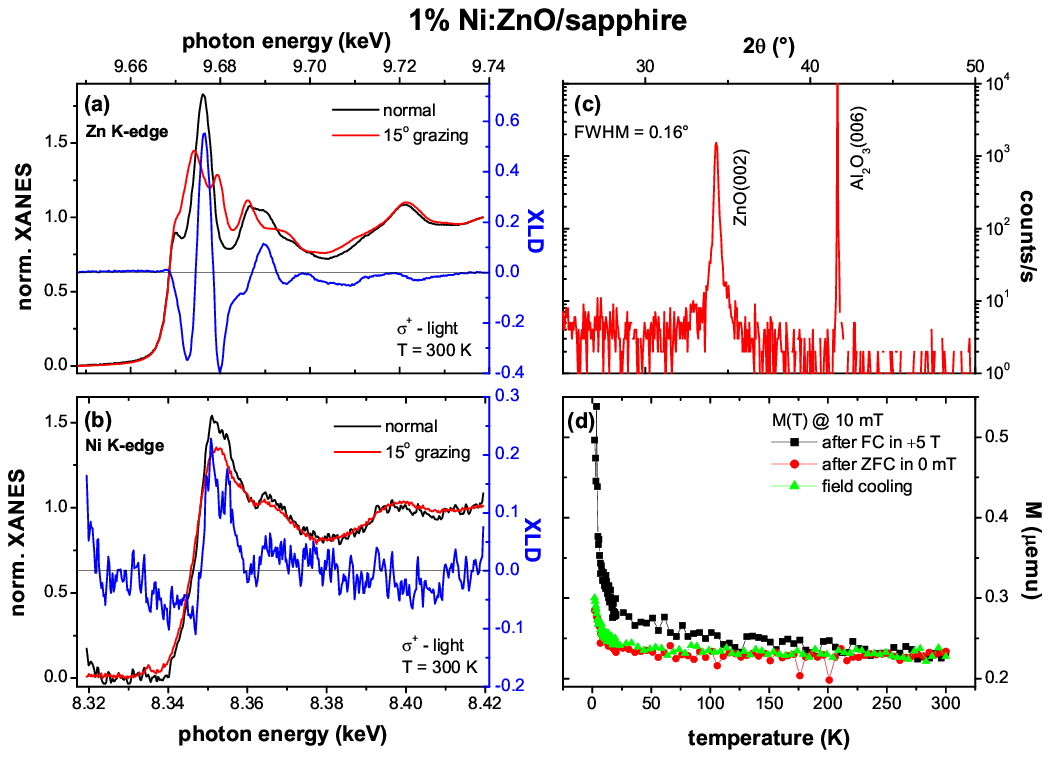}}
\caption{XANES and XLD spectra of 1\% Ni:ZnO at the (a) Zn and (b) Ni $K$-edges. Structural characterization by (c) XRD and (d) $M(T)$ curves measured with SQUID magnetometry. \label{fig3}}
\end{figure}

In a next step it was tried to increase the Ni doping level beyond the impurity limit of the Ni-diffused single crystals. For that purpose, ZnO epitaxial films have been grown on c-plane sapphire substrates using RMS. First, a very low Ni-concentration of nominally 1\% was tried using co-sputtering. Figure \ref{fig3} shows the XANES and XLD spectra at the Zn (a) and Ni (b) $K$-edges. The total amplitude at the Zn $K$-edge of 0.93 is slightly smaller than the optimum value of 1.15 for a ZnO reference sample reported earlier \cite{NKN11}. At the Ni $K$-edge in Fig.\ \ref{fig3} (b) a lot of noise is visible which is on the one hand due to the still rather low amount of Ni in this thin epitaxial film. On the other hand, also here the XLD had to be derived by grazing vs. normal XANES measurements so that further background issues arise. Therefore, not much quantitative information can be drawn from the size and shape of the XLD. Nonetheless it is visible that the overall spectral shape is somewhat similar to the simulation in Fig.\ \ref{fig1}(b), so that there is at least some indication that Ni may also sit on substitutional Zn sites and the overall amplitude of the XLD of about $-0.1$ to $0.2$ is rather similar to the Ni diffused sample shown in Fig.\ \ref{fig1}(d).  

The good crystalline structure of the 1\% Ni:ZnO film is further corroborated by an XRD reflection corresponding to the (002) peak of the ZnO wurtzite lattice shown in Fig.\ \ref{fig3}(c) with a rather narrow full-width at half maximum (FWHM) of only 0.19$^{\circ}$, i.\,e., almost as low as typical Co:ZnO films of highest crystalline quality which are below 0.15$^{\circ}$ \cite{NOK10}; however the onset of a small shoulder in the XRD peak at higher angles is visible. Taking all structural information together the 1\% Ni:ZnO film is in the wurtzite structure with a reasonable crystallinity and at least a fraction of the Ni occupies Zn lattice sites and no direct evidence for secondary phases can be found by means of XRD or at the Zn $K$-edge; however, the quality of the Ni $K$-edge spectra does not allow to rule out the presence of secondary Ni phases such as metallic Ni and/or, e.\,g., rocksalt NiO and the shoulder in the XRD also hints in that direction. Here SQUID magnetometry can give further evidence of eventual secondary phases and the $M(T)$ curves under FC and ZFC conditions as well as while cooling down are shown in Fig.\ \ref{fig3}(d). While no significant separation between the $M(T)$- curve after ZFC (red circles) and the $M(T)$ curve while cooling down (green triangles) is visible within the noise level, the $M(T)$ curve after FC in 5~T (black squares) shows a small separation up to around 150~K to 200~K which may indicate the presence of a small fraction of other magnetic phases, presumably antiferromagnetic ones. We will come back to this point later-on. Overall, the dominant magnetic contribution in the $M(T)$ curves appears to be paramagnetic and the absence of ferromagnetism can be safely ruled out, i.\,e., the presence of a ferromagnetic secondary phase due to metallic Ni can be excluded. 

\begin{figure}[tb]
\resizebox{1\columnwidth}{!}{\includegraphics{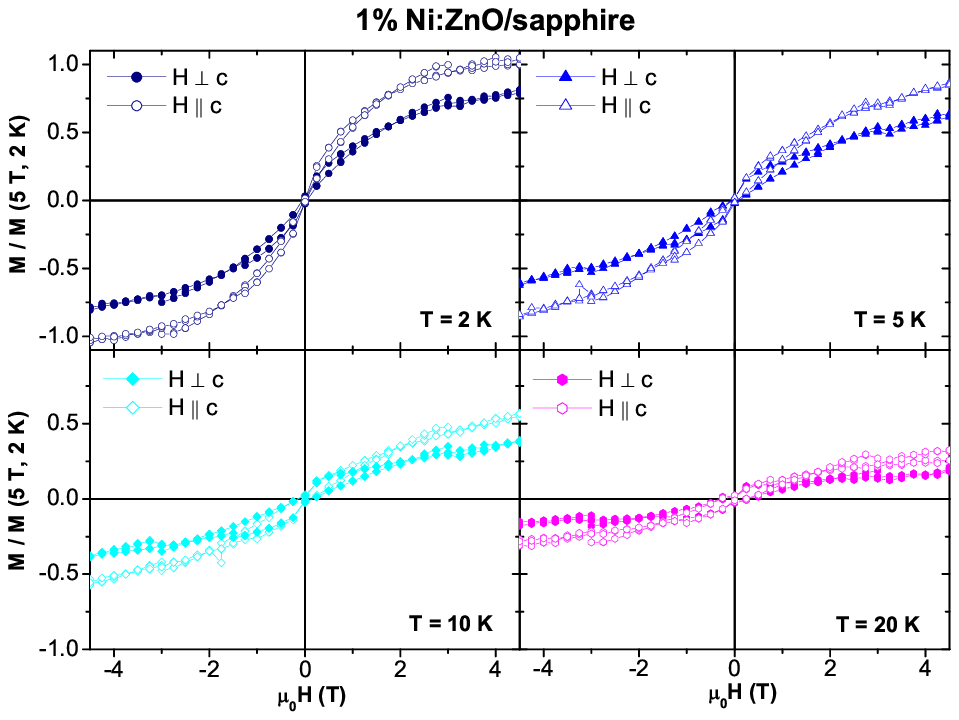}}
\caption{$M(H)$ curves of 1\% Ni:ZnO measured while applying the magnetic field in the film plane (full symbols) as well as out-of-plane (open symbols) at four different temperatures from 2~K to 20~K.  \label{fig4}}
\end{figure}

Figure \ref{fig4} shows $M(H)$ curves of the 1\% Ni:ZnO film at 2~K, 5~K, 10~K and 20~K where the external magnetic field was applied either in the film plane, i.\,e., $H \perp c$ (full symbols), or out-of-plane, i.\,e., $H || c$ (open symbols). For better comparison, the data were normalized to $M$ recorded at 2~K and 5~T for the out-of-plane geometry, which is the largest magnetic response of the sample. A clear anisotropy is visible despite the rather high noise level stemming from the overall small magnetic signal. Similar to the Ni-diffused sample shown in Fig.\ \ref{fig1}(c) the anisotropic paramagnetism has an easy $c$ axis and a hard $a$ plane being indicative of at least a significant fraction of Ni being substitutionally incorporated on Zn lattice sites. Note, that no ESR line of Ni$^{3+}$ could be detected so that neither the XANES nor the ESR can give a direct experimental evidence for the oxidation state, which can be assumed to be 2+. The temperature dependence of the anisotropic $M(H)$ curves in Fig.\ \ref{fig4} demonstrates that within the noise level the anisotropy is still visible at 10~K and has vanished at 20~K which is rather similar to the temperature dependence of the single ion anisotropy of Co$^{2+}$ in ZnO \cite{NKO10}, again underlining the presence of a significant amount of Ni on Zn lattice sites. However, to draw conclusions on a more quantitative ground the Ni content has to be increased to increase the attainable signals in XANES, XLD and SQUID.

\begin{figure}[tb]
\resizebox{1\columnwidth}{!}{\includegraphics{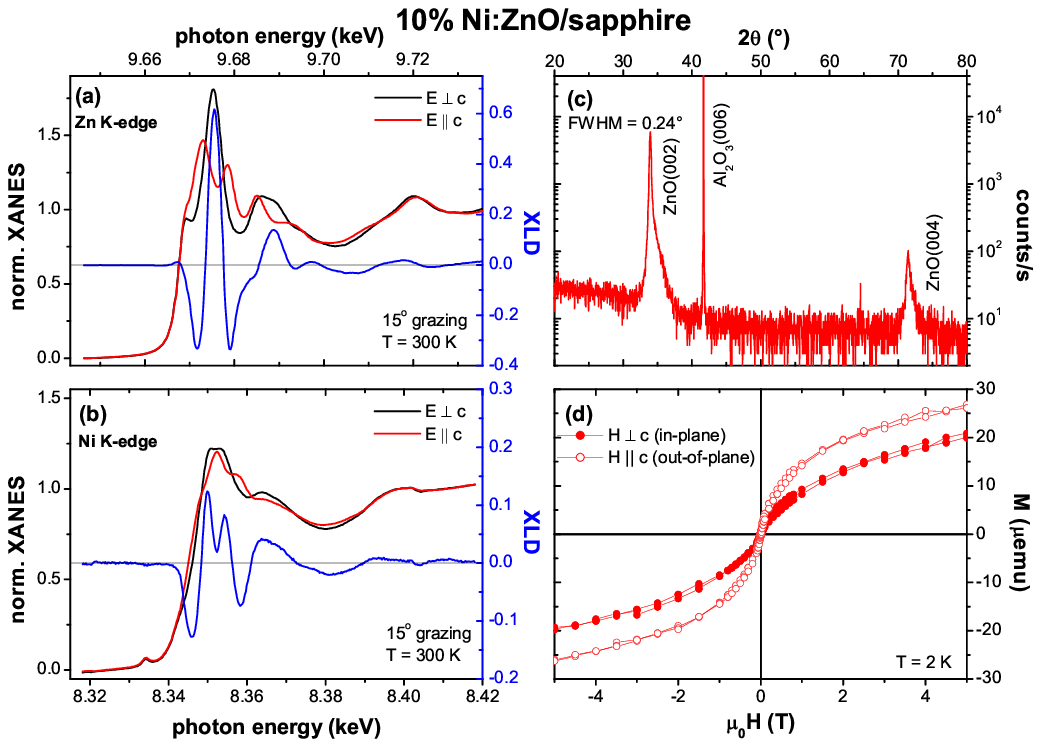}}
\caption{XANES and XLD spectra of 10\% Ni:ZnO measured at the (a) Zn and the (b) Ni $K$-edges. Structural characterization by (c) XRD and (d) $M(H)$ curves at 2~K measured while applying the magnetic field in the film plane (full symbols) as well as out-of-plane (open symbols).\label{fig5}}
\end{figure}

Figure \ref{fig5} shows the results for the 10\% Ni:ZnO sample with the highest attainable structural quality which was grown from a Ni/Zn-10/90 composite target. Similar to the 1\% Ni:ZnO sample the overall XLD signal at the Zn $K$-edge shown in Fig.\ \ref{fig5}(a) is found to be 0.94, while at the Ni $K$-edge displayed in Fig.\ \ref{fig5}(b) the spectral shape of the XLD of the Ni-diffused sample in Fig.\ \ref{fig1}(d) is nicely reproduced; however, this time the quarter-wave plate was available and thus the signal to noise ratio has been significantly improved and the background is nicely suppressed. Nevertheless, the overall size and shape of the experimental XLD significantly deviates from the FDMNES simulation in Fig.\ \ref{fig1}(b). The fact that in the case of Co the agreement between experiment and theory had been remarkable, this discrepancy can have two potential reasons: (i) the incorporation of the Ni is not exactly at the lattice position of the substituted Zn ion, or (ii) only part of the Ni is substitutionally incorporated. In a naive picture (i) should alter the spectral shape while (ii) would reduce the overall amplitude, although in reality both effects will be intertwined. For the purpose of the present work it is however sufficient to conclude that eventhough not all Ni ions are incorporated on ideal Zn lattice sites, at least a significant fraction is. This can be further refined by the XRD shown in Fig.\ \ref{fig5}(c) where the wide-angle $\omega - 2\theta$ scan shows no other reflection beyond the ZnO(002), ZnO(004) and the Al$_2$O$_3$(006) peaks but the FWHM is increased to 0.24$^{\circ}$ compared to the 1\% Ni:ZnO sample and the shoulder towards higher angles at the ZnO peak is now clearly visible, indicating the onset of some structural degradation. On the other hand, the $M(H)$-curves in Fig.\ \ref{fig5}(d) demonstrate that the single ion anisotropy is still existing and its sign and size does not change in comparison with the 1\% Ni:ZnO sample. In other words, also the 10\% Ni:ZnO sample shows all the characteristics of the predominant occurrence of Ni$^{2+}$ substitutionally incorporated on Zn lattice sites similar to the Ni-diffused bulk ZnO as well as the 1\% Ni:ZnO epitaxial film. Note, that all sputtered Ni:ZnO films turned out to be highly resistive.

To summarize this part, despite secondary phases cannot be fully ruled out, it appears very likely that Ni is predominantly incorporated into the ZnO host lattice in its formal 2+ oxidation state on tetrahedrally coordinated Zn lattice sites based on the similarities between the XANES and XLD at the Ni $K$-edge. While Ni diffusion has led to significant changes in the conductivity, the Ni incorporation during sputtering did not, underlining the indirect role of Ni in the observed increase in conductivity in the in-diffused bulk samples. However, the magnetic properties remain unchanged exhibiting paramagnetism with a significant single-ion anisotropy at low temperatures which has a comparable size but opposite sign compared to Co$^{2+}$ in ZnO. These findings allow to address a new aspect in these types of samples. It has been shown, that the next cation neighbor interaction of Co in ZnO is antiferromagnetic, leading to a significant reduction of the effective magnetic moment per Co atom with increasing Co concentration, see, e.\,g., \cite{NHL16} and Refs. there-in. Ni$^{2+}$, which is a 3$d^8$ ion and Co$^{2+}$, which is a 3$d^7$ ion are expected to have a different magnetic moment, so that co-doping of both ions into ZnO may be a path to reduce the magnetic compensation due to a statistical incorporation of both Ni and Co, even if all respective next cation neighbor couplings are antiparallel. This hypothesis will be addressed in the following section.

\subsection{Wurtzite Ni- and Co-codoped ZnO epitaxial films}

\begin{figure}[tb]
\resizebox{1\columnwidth}{!}{\includegraphics{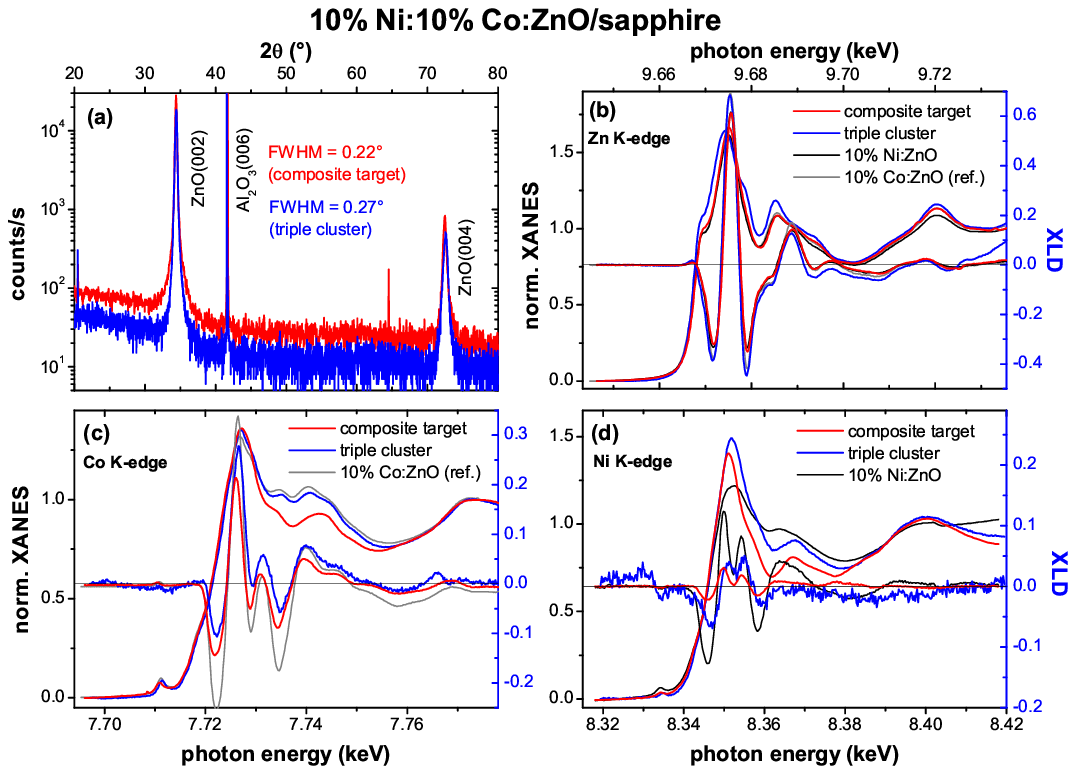}}
\caption{10\% Co / 10\% Ni co-doped ZnO films grown with different sputtering setups and targets. (a) shows the structural properties by means of XRD, while the XANES as well as XLD spectra are displayed for (b) the Zn, (c) the Co and (d) for the Ni $K$-edges together with reference spectra for either Ni- or Co-doped ZnO. \label{fig6}}
\end{figure}

The starting point of testing the hypothesis of a partial lifting of the magnetic compensation due to next cation neighbor antiferromagnetic interactions is a co-doping of wurtzite ZnO with 10\% Co and 10\% Ni by sputtering from a Ni/Co/Zn-10/10/80 composite target. The total amount of magnetic dopants is thus 20\% which is right at the coalescence limit of the hcp cationic sublattice. This should avoid coalescence-induced long range magnetic order like reported in \cite{NHL16}. To allow for a larger variation of the respective Ni and Co content, a sample with nominally the same composition has been fabricated in a triple cluster system. Figure \ref{fig6}(a) shows the XRD diffractograms of the two samples with the best attainable crystalline quality, respectively. Both XRD are devoid of any signs of secondary phases; the FWHM of 0.22$^{\circ}$ and 0.26$^{\circ}$ is roughly similar to the 10\% Ni:ZnO. This can be compared to attempts to achieve high quality 20\%-doped ZnO with only one atomic species. 20\% Co:ZnO results in a clearly higher crystalline quality with a FWHM of only 0.14$^{\circ}$ \cite{NHL16}. In contrast all attempts to grow a 20\% Ni:ZnO sample with reasonable quality by sputtering from a Ni/Zn-20/80 composite target have failed and the smallest attainable FWHM was 0.4$^{\circ}$ (not shown). Obviously, an increased Ni content in ZnO reduces the structural quality thus indicating the formation of unwanted secondary phases. In other words, the solubility of Ni in ZnO is limited, corroborating earlier reports of the onset of phase separation around 5\% \cite{GoS13}. Therefore, the tetrahedral coordination of Ni appears to be less favorable in ZnO compared to Co where cationic substitution up to 60\% are attainable \cite{NHL16}.

Turning back to the 10\% Ni- 10\% Co-codoped ZnO samples, in Fig.\ \ref{fig6} the XANES and XLD spectra of both samples measured at the (b) Zn, (c) Co, and (d) Ni $K$-edges are compared to the respective reference spectra of the 10\% Ni:ZnO sample from Fig.\ \ref{fig5} and the 10\% Co:ZnO sample from \cite{NOK10}. It should be noted, that the triple-cluster sample was again only measured by grazing vs. normal incidence while all other could be measured using the quarter-wave plate. Therefore, the quantitative conclusions for the triple-cluster samples are somewhat limited, as can be seen from the obvious background issue at the Zn $K$-edge or the reduced signal to noise ratio at the Ni and Co $K$-edges. In turn, it nicely illustrates how important the use of a quarter-wave plate is for recording meaningful high-quality XLD spectra. It is obvious from the size of the XLD at the Zn $K$-edge, that the two co-doped samples can match the good overall crystalline quality of the ZnO host lattice of both reference samples at a local scale thus corroborating the XRD findings. In contrast, at the Co $K$-edge the size of the XLD signal is reduced by almost a factor of 2 compared to the ideal 10\% Co:ZnO. This indicated that the Ni co-doping also leads to a reduced local structural quality for the Co dopant. Similarly, the XLD signal at the Ni $K$-edge is strongly reduced in amplitude compared to the 10\% Ni:ZnO while the spectral shape remains unchanged. Thus, a slight reduction in the overall structural quality of the ZnO host crystal is evidenced by the globally sensitive XRD as well as the XLD, which measures the structural quality on a local scale. This goes hand in and with a reduced amount of substitutionally incorporated Co and Ni ions in the co-doped samples compared to the singly doped ones.

\begin{figure}[tb]
\resizebox{1\columnwidth}{!}{\includegraphics{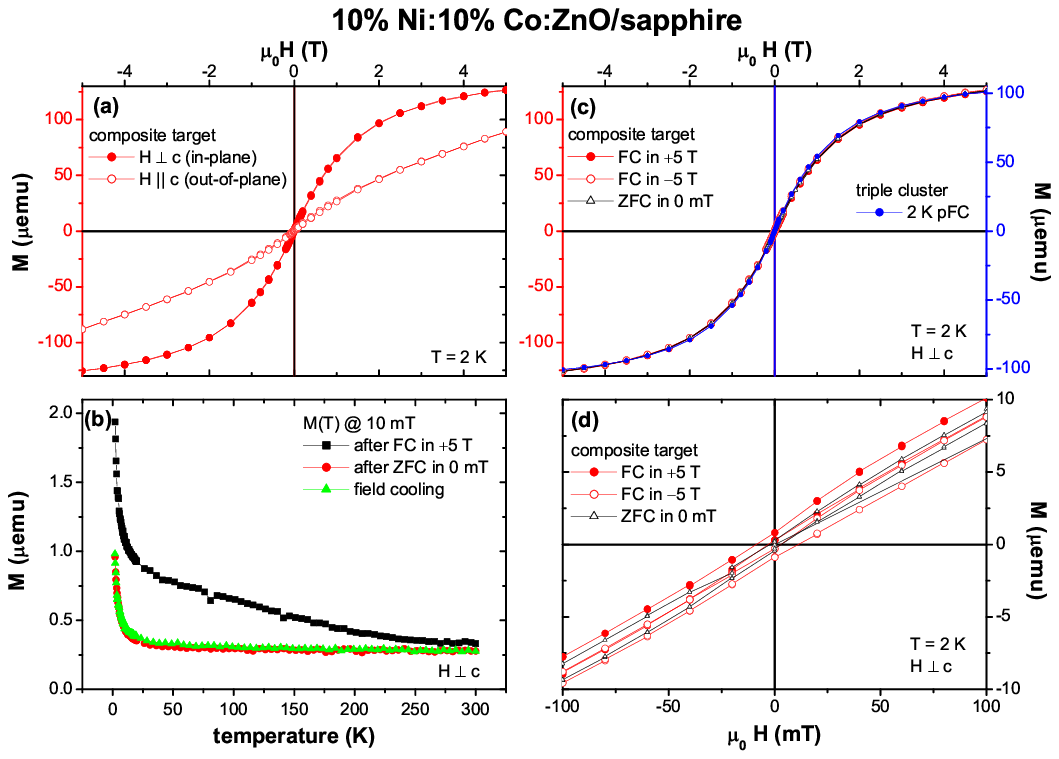}}
\caption{(a) $M(H)$ and (b) $M(T)$ curves for the 10\% Co / 10\% Ni co-doped ZnO film grown from the composite target. (c) comparison of the $M(H)$ curves at 2~K between the two samples as well as cooling-field dependence; the low field region is magnified in (d).  \label{fig7}}
\end{figure}

The simultaneous incorporation of Co and Ni has also consequences for the magnetic properties. Figure \ref{fig7}(a) shows the $M(H)$ curves measured at 2~K with the magnetic field applied in- (full symbols) and out-of-plane (open symbols). A pronounced anisotropy is visible which is however opposite to that of 10\% Ni:ZnO. It merely resembles the single ion anisotropy of Co:ZnO and the increased curvature is a first hint towards an increased effective magnetic moment. The $M(T)$ curves are plotted in Fig.\ \ref{fig7}(b) under FC and ZFC conditions. No significant separation between the ZFC curve and the curve during cool down is visible, indicating the absence of a (ferro)magnetic hysteresis. On the other hand, the FH curve significantly deviates from the others up to about 250~K. Since it is only present after FC in 5~T, it can be explained by a field imprinted magnetization of predominantly antiferromagnetic objects. Figure \ref{fig7}(c) therefore compiles the in-plane $M(H)$ curves at 2~K after FC in $+5$~T, $-5$~T and in nominally zero field. These types of measurements have previously been used to study the field imprinted magnetization of uncompensated antiferromagnetism in highly doped Co:ZnO \cite{HNS16}. On the full field scale the three curves are indistinguishable and also the comparison between the samples from the composite target and the triple cluster are identical. Only in the low field region shown in Fig.\ \ref{fig7}(d) a slight vertical shift of the $M(H)$ curves after FC is visible, which reverses upon reversing the cooling field direction. However, virtually no hysteresis is visible. Nevertheless, the small difference between the $M(H)$ curves between FC and ZFC is the source of the deviation between FH and ZFC in the $M(T)$ curves in Fig.\ \ref{fig7}(b). Following the model introduced in \cite{HNS16} for Co:ZnO one has to adopt it to the anhysteretic case like done for Cu:ZnO \cite{NVW19}. Assuming that small precipitates of antiferromagnetic NiO are formed which are too small to be blocked, one would have uncompensated magnetic moments at the surface of these precipitates which give rise to the tiny field-imprinted magnetization. The hypothesis of small (of the order of few nm) NiO precipitates could explain the observed integral magnetic properties and furthermore is consistent with the strongly reduced substitutional incorporation of Ni as evidenced by XLD. Also the missing evidence for secondary phases within the limits of XRD are in favor of the formation of small precipitates. To this magnetic contribution of at least part of the Ni forming the NiO precipitates the contributions of the substitutionally incorporated Ni and Co with their respective single ion anisotropy is added up. The fact that the overall anisotropy is dominated by the behavior of the Co is also supportive of a better substitutional incorporation of Co compared to Ni and thus consistent with the findings of the XLD in Fig.\ \ref{fig6}. 

\begin{figure}[tb]
\resizebox{0.7\columnwidth}{!}{\includegraphics{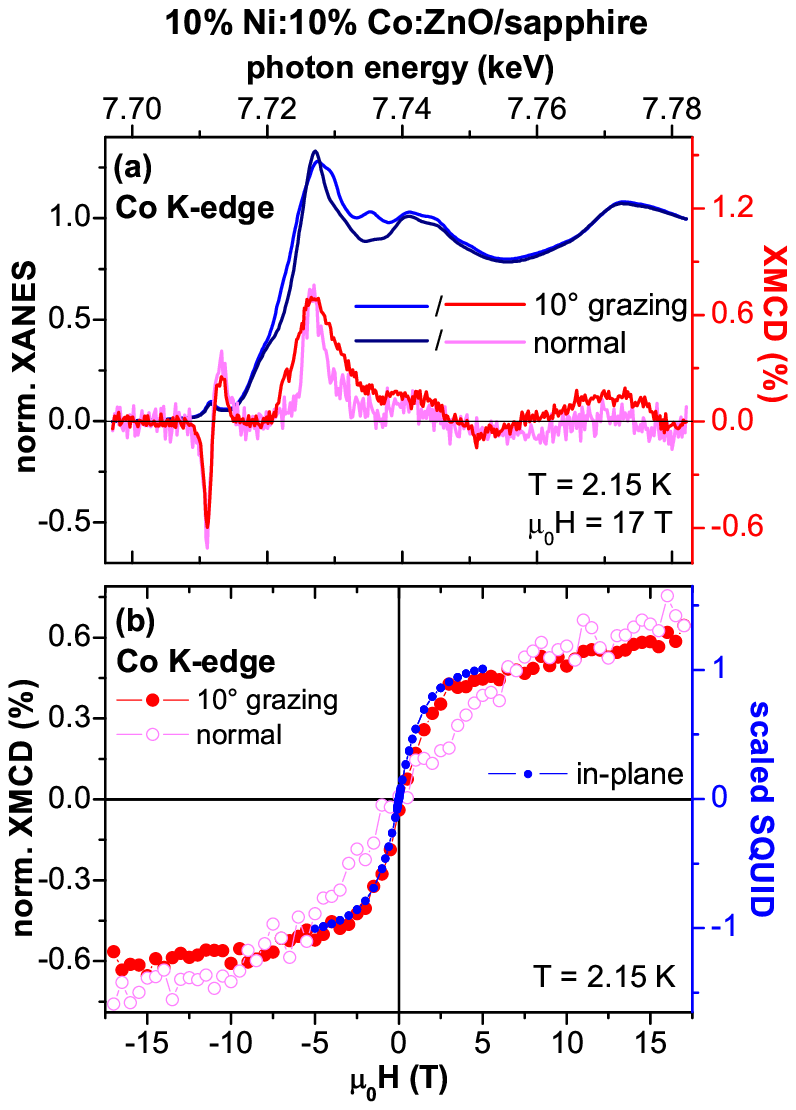}}
\caption{(a) XANES and XMCD spectra of the 10\% Co / 10\% Ni co-doped ZnO film for grazing and normal incidence recorded at the Co $K$-edge. (b) XMCD$(H)$ curves for in- and out-of-plane in comparison to the scaled $M(H)$ curves measured by SQUID. \label{fig7a}}
\end{figure}

In an attempt to investigate the magnetic coupling between Co and Ni in the 10\% Co: 10\% Ni:ZnO sample, element-specific magnetometry by means of XMCD has been carried out at the Ni and Co $K$-edges, respectively. However the size of the XMCD signal at the Ni $K$-edge was too low to be reliably detected; not even the relative sign with respect to the Co $K$-edge could be detected so that no information on the relative orientation of the Co and Ni magnetic moments is available. Figure \ref{fig7a}(a) shows the XANES and XMCD spectra at the Co $K$-edge recorded at 2~K and 17~T for normal and grazing incidence. The typical spectral shape of the XMCD for Co$^{2+}$ in Co:ZnO is observed, in particular the characteristic feature for metallic Co is absent, see \cite{NOK10}. At the maximum XMCD signal also XMCD$(H)$ curves have been recorded in both orientations and the result ist shown in Fig.\ \ref{fig7a}(b) together with the scaled SQUID data from Fig.\ \ref{fig7} for the in-plane orientation. The match of the functional behavior is rather good so that the overall shape of the $M(H)$ curve of the integral SQUID magnetometry can be readily explained by the observed Co behavior as measured with elemental selectivity at the Co $K$-edge. Likewise, the identical single-ion anisotropy as observed using SQUID in Fig.\ \ref{fig7}(a) is also clearly visible in the XMCD$(H)$ behavior at the Co $K$-edge in Fig.\ \ref{fig7a}(b), however, due to the increased noise level a direct comparison between XMCD and SQUID is not meaningful. Nevertheless, it is obvious that the integral magnetic behavior is determined by the behavior of the Co.

There is one interesting aspect in the XMCD at the Co $K$-edge in Fig.\ \ref{fig7a}(a) which should be discussed in more detail. The maximum XMCD signal is 0.6\% for a nominal Co content of 10\% in the Ni/Co-codoped sample. In contrast, in 10\% Co:ZnO samples the XMCD at 2~K and 17~T is 0.4\%, for 20\% Co:ZnO 0.3\% \cite{NNW12,NHL16}. The reduction of the effective magnetic moment with increasing Co content is attributed to the magnetic compensation due to antiferromagnetic next cation neighbor interactions. In turn, the increase of the XMCD signal upon Ni codoping for the identical Co concentration indicates that the magnetic compensation is partially lifted due to the Ni codoping. In other words, assuming that also Co/Ni next cation neighbors couple antiferromagnetically, some fraction of the Ni, which has to be located on Zn lattice sites in-between Co ions gives rise to an increased effective magnetic moment of the Co by forming Co-O-Ni-O-Co-\dots configurations, where Co effectively couples parallel, while the Ni, which has a smaller magnetic moment than Co, is aligned antiparallel to the Co moments. Naturally, in tetrahedral coordination as in wurtzite ZnO the real situation is more complex and some degree of frustration and presumably non-collinear arrangements will come into play; however, the net increase of the effective Co moment upon Ni co-doping remains an experimental fact and in-turn suggests that at least part of the Ni is still incorporated on Zn lattice sites as also corroborated by the remaining small XLD signal in Fig.\ \ref{fig6}(d). Unfortunately, this cannot be further confirmed by increasing either the Ni or the Co content, or both, because the structural quality of the samples starts to degrade so that meaningful conclusions cannot be drawn anymore. This limited solubility of Ni in ZnO corroborates earlier findings \cite{GoS13} also for the case of Co codoping.

To summarize this part, in epitaxially grown Ni/Co-codoped ZnO the Ni is still incorporated as Ni$^{2+}$ mostly on substitutional Zn lattice sites. The magnetic properties are however dominated by the Co single ion anisotropy and the substitutional Ni leads to a partial lifting on the magnetic compensation of the Co evidenced by an increased effective magnetic moment of the Co. Nevertheless, Ni starts to form secondary phases, which can only be evidenced indirectly via the reduced XLD signal and the onset of a field-imprinted magnetization. The most likely secondary phase are tiny NiO precipitates, which go undetected in XRD, and have no specific XLD signature due to their cubic structure. The Ni is still in its formal 2+ oxidation state thus leaving the XANES virtually unchanged. The only open point is the temperature at which the field imprinted magnetization vanishes, about 250~K, as seen in Fig.\ \ref{fig7}(b), is too low for the bulk N\'eel-temperature of 523\,K for NiO \cite{Rot58}, which could be an indication that also some small fraction of rocksalt CoO is formed which would have a bulk N\'eel-temperature of 291\,K \cite{Rot58}. Of course also the formation of (Ni/Co)O as reported in \cite{MSP19} is possible which would have a N\'eel-temperature between CoO and NiO. On the other hand, finite size effects may also come into play so that a clear identification of the phase giving rise to the field imprinted magnetization cannot be unambiguously identified. Concerning the incorporation and valence of Ni in wurtzite ZnO the dominating species remains Ni$^{2+}_{Td}$, especially at low dopant concentrations, while at higher Ni concentrations secondary phase formation, presumably NiO starts to set in. Higher Ni concentrations will be discussed in the following for the case of Ni doping of the ZnCo$_2$O$_4$ spinel.      

\subsection{Ni-doped ZnCo$_2$O$_4$ spinel}

The ZnCo$_2$O$_4$ spinel fabricated by RMS has already been reported with its characteristic signatures in the Zn and Co $K$-edge XANES and XLD \cite{HNO15}. These spectra can serve as reference for the Ni-codoping series which will be discussed in this section. It is the aim to elucidate the cationic occupation of Zn, Ni and Co along with their respective valence state. Although it is common sense, that Zn$^{2+}_{Th}$ is the predominant species for Zn in spinels, we will also consider this cation along with the two others. The actual Ni concentrations of this series from 5~W to 35~W Ni is determined by the individual sputter rates and only a rough estimate can be provided. If one just refers to the cationic sublattice, the ideal distribution between Zn and Co would be 1:2 in the spinel, i.\,e., 33\% Zn, but we have already provided evidence by Rutherford backscattering spectrometry (RBS) that these films are typically Zn rich with a Zn content corresponding to close to 40\% \cite{DAC21}. Since the sputter rates of Co and Ni are rather similar, the 5~W Ni sample should contain around 9\% Ni with respect to Co, i.\,e., $5-6$ cationic-\%. The 20~W Ni sample then would have roughly $20-25$\% while the relative Zn content will be slightly reduced towards the ideal spinel one. The 35~W Ni sample on the other hand would have clearly reduced Zn content around 20 cationic-\% and Ni will be of the same order. However, the exact composition is not decisive for the following observations. 

\begin{figure}[tb]
\resizebox{1\columnwidth}{!}{\includegraphics{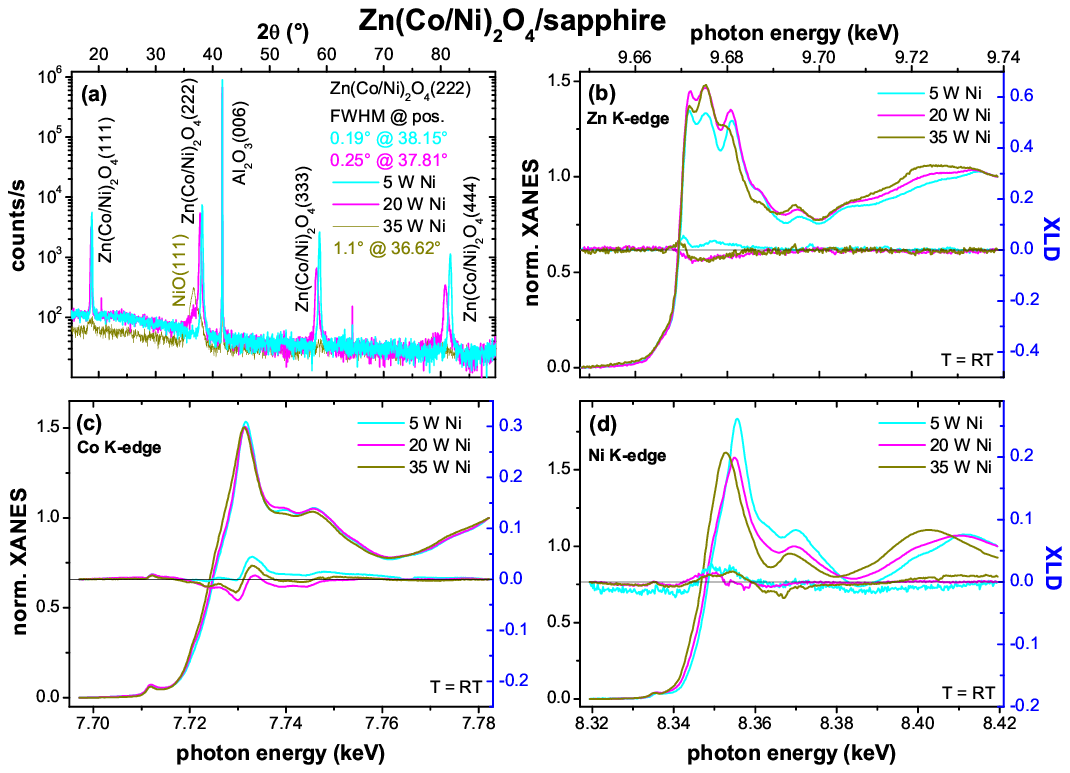}}
\caption{(a) XRD diffractograms of ZnCo$_2$O$_4$ spinel with three different Ni concentrations. The resulting XANES, and XLD spectra are shown for the (b) Zn, (c) Co and (d) Ni $K$-edges. \label{fig8}}
\end{figure}

Figure \ref{fig8}(a) shows the XRD of the three samples grown with different sputter power of Ni from 5~W to 35~W. The 5~W samples shows clear spinel-(111) reflections and higher orders with a low FWHM of only 0.19$^{\circ}$, which indicates good crystalline quality matching or even exceeding those of the undoped ZnCo$_2$O$_4$ spinel. The changes for the 20~W Ni sample are rather weak with a slight shift to lower angles and a slightly increased FWHM of 0.25$^{\circ}$. However, a small shoulder at lower angles is already recognizable, in particular at the (444) peak. In contrast, the 35~W Ni sample shows hardly any (111) spinel peaks but a broad, rather clear peak at 36.62$^{\circ}$ which can be associated with NiO(111). We take this as a first direct indication that NiO is formed as secondary phase. The corresponding XANES and XLD spectra are shown in Fig.\ \ref{fig8} for the (b) Zn, (c) Co and (d) Ni $K$-edges, respectively. With the exception of Co no significant XLD signal can be observed. The tiny XLD at the Co $K$-edge was already observed in the undoped ZnCo$_2$O$_4$ spinel \cite{HNO15} and will be disregarded in the following, since no conclusive assignment of this XLD signature to a certain Co species could be made. Overall, the Co $K$-edge does not change significantly over the Ni codoping-series, whereas both other edges, Zn and Ni, show pronounced changes. While the Ni $K$-edge for the 5~W and 20~W is more indicative for Ni$^{3+}$, there is a clear shift of the Ni $K$-edge towards lower energies for the 35~W sample. Together with the NiO peak seen in XRD we take this as indication that at least a substantial fraction of Ni is present as Ni$^{2+}$, presumably as cubic NiO. Therefore, we will exclude this sample from a further in-depth analysis of (local) structure and valence.    

\begin{figure}[tb]
\resizebox{1\columnwidth}{!}{\includegraphics{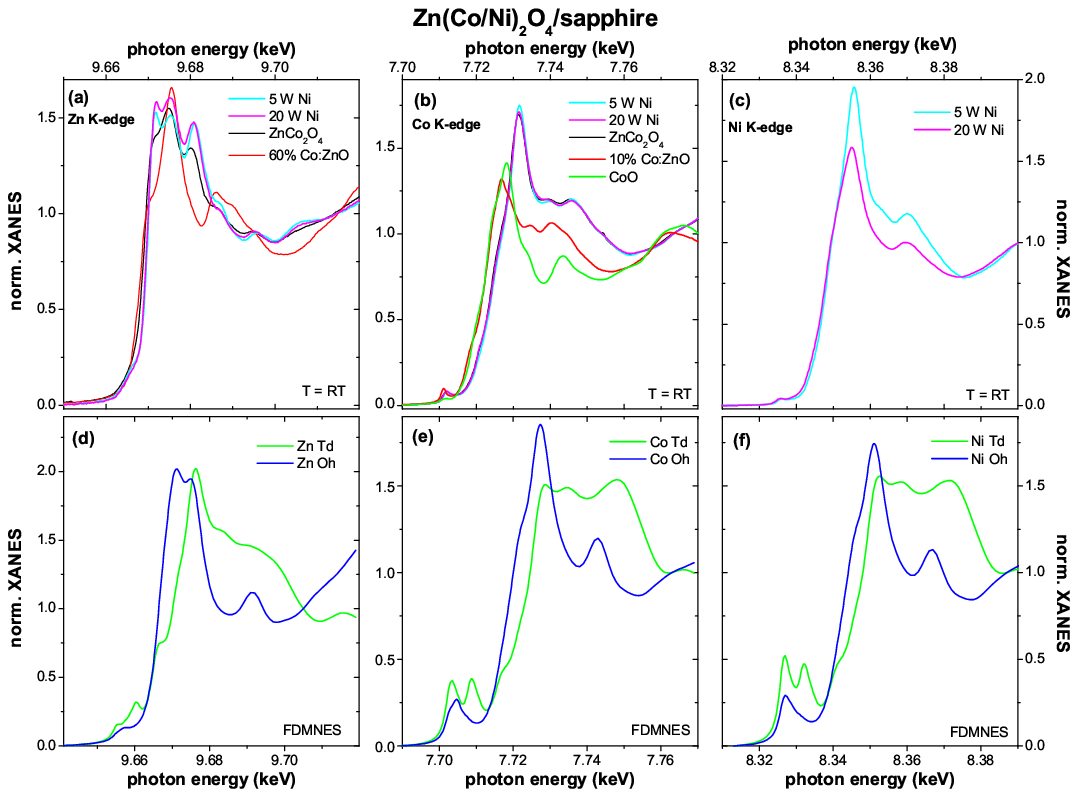}}
\caption{Comparison of the experimental XANES of the Ni-doped ZnCo$_2$O$_4$ spinel with experimental references spectra (top) and FDMNES simulations (bottom) for the absorbing atom located on tetrahedral (Td) or octahedral (Oh) lattice sites for the Zn (a/d), the Co (b/e), and the Ni (c/f) $K$-edges, respectively.  \label{fig9}}
\end{figure}

In Fig.\ \ref{fig9} the two Ni-codoped samples are shown in comparison to reference spectra for wurtzite 60\% Co:ZnO and the ZnCo$_2$O$_4$ spinel for the (a) Zn, (b) Co and (c) Ni $K$-edges, respectively. These are compared with respective simulations using the FDMNES code \cite{BuJ09}. These simulations were carried out using an ideal spinel lattice without distortions of the O-octahedra and a cubic lattice constant of 0.80946~nm. In general, the tetrahedral A sites were populated with Zn and the octahedral B sites with Co. This way Zn$_{Td}$ and Co$_{Oh}$ were simulated. To check for possible A/B disorder, either a pair of Zn/Co cations were exchanged, or one tetrahedral Zn was replaced with Co, or an octahedral Co was replaced with Zn. This way Zn$_{Oh}$ and Co$_{Td}$ were simulated. Both alternatives do not lead to significant changes in the result and Fig. \ref{fig9} only shows the version with an additional Zn or Co atom. This approach was also used for the Ni XANES, where either one Zn$_{Zd}$ or one Co$_{Oh}$ was replaced with Ni. Furthermore, for all simulations it was checked, that the overall appearance of the XANES (there was no XLD since the input structure was ideally cubic) is identical, whether the calculation was done using self-consistent potentials (SCF-option of the FDMNES code) or not. Likewise also the FDM method \cite{GGS15} was compared with the multiple scattering variant (green-option in the FDMNES code). Again, the overall structure of the simulated XANES were rather similar. The most noticeable changes were seen at the pre-edge feature of the $K$-edges, which is however too strong in any of the simulations compared to the experiment. Since the non-selfconsistent muffin-tin potentials with the multiple scattering calculations are the fastest, we are only showing those in Fig.\ \ref{fig9}. This variant has already given surprisingly good simulations of the experimental $K$-edge spectra for wurtzite Co:ZnO \cite{NOY08} but in the present case the agreement between experiment and simulation is less good; however, none of the other variants has led to a noticeable improvement. Note, that we did not try to optimize the input structure because we do not have any independent experimental information with regard to local distortions etc. so that there would be too many parameters to be optimized independently which in turn would be difficult in cases where the agreement between experiment and simulation is not very good. Therefore, we will use mainly the basic structure of the simulated XANES complemented by a comparison with reference spectra where possible. 

First, we turn to the experimental XANES of the Co $K$-edge in Fig.\ \ref{fig9}(b) which is compared with reference spectra of the undoped spinel, wurtzite 10\% Co:ZnO and cubic CoO. There is a clear discrepancy between latter two, where Co is in its formal 2+ oxidation state, but the spectra are virtually indistinguishable from the undoped spinel. Therefore, one can conclude that Co only exists in its 3+ formal oxidation state as expected for ideal ZnCo$_2$O$_4$ which is a normal spinel. However, if one looks closer to the XANES, one notices, that there is a small double peak after the large peak of the main absorption, which is absent in the simulations for Co$_{Oh}$ in Fig.\ \ref{fig9}(e). In contrast, the simulation of Co$_{Td}$ exhibits no marked main peak but a triple-peak structure of roughly the same intensity. Comparing experiment with simulation the only way of getting the small second peak in the Co $K$-edge XANES is to assume that a small fraction of Co$_{Td}$ is formed but the dominant species is Co$^{3+}_{Oh}$. This is not a very surprising result, because also the Co$_3$O$_4$ spinel exists, where also Co$_{Td}$ has to exist. Turning to the simulations of the Ni $K$-edges in Fig.\ \ref{fig9}(f) one notices, that both, tetrahedral and octahedral species are rather similar for Ni as for Co. However, the second peak after the main absorption is almost absent in the experimental XANES for the 5~W Ni sample in Fig.\ \ref{fig9}(c). Thus, only a tiny amount if any of the Ni is incorporated on a tetrahedral site and the predominant species is Ni$_{Oh}$ presumably in its 3+ oxidation state. The finding is in line with the fact that both NiCo$_2$O$_4$ and NiFe$_2$O$_4$ are known to be inverse spinels, i.\,e., also there Ni clearly favors octahedral sites. For the 20~W Ni sample a clear reduction in the overall intensity of the XANES occurs and the peak which is indicative of tetrahedral sites has completely vanished. In addition, no shift of the onset of the edge is visible. Therefore, one cannot infer any significant alteration of the prevailing Ni species for the spinel. However, the onset of a shoulder close to the position of the NiO reflex in the XRD for this sample may be interpreted as very first indication of phase separation which would also alter the Ni $K$-edge accordingly. In fact, the overall appearance of the Ni $K$-edge of the 20~W sample in Fig.\ \ref{fig8}(d) is in-between those of the 5~W and 35~W samples.

In a second step we turn to the Zn $K$-edges where also some changes are already visible in Fig.\ \ref{fig8}(b) where three main peaks of varying intensity can be seen. The simulations of Zn$_{Td}$ and Zn$_{Oh}$ in Fig.\ \ref{fig9}(d) are quite different from those of Co or Ni: here the tetrahedral site results in one clear main peak, while the octahedral site shows a double-peak. Obviously, none of the two simulations matches the experimental XANES with its three peaks but it appears to be reasonable to assume that Zn$_{Td}$ alone would never lead to a better match. To get more independent experimental evidence the Zn $K$-edge of the two Ni-codoped spinels is compared with the undoped ZnCo$_2$O$_4$ spinel as well as with the 60\% Co:ZnO film which were both grown from the same target \cite{HNO15}. The wurtzite Co:ZnO, where only tetrahedral sites exist, clearly shows a single main absorption peak and thus can be taken as a characteristic spectroscopic signature of Zn$_{Td}$. Of course, the local distances in wurtzite vs. spinel are different, but also the simulations of the spinel give a comparable result with a single peak. For the undoped spinel, two additional peaks evolve left and right from the main absorption. For comparison with the simulations the first peak may be problematic, because it is close to the rising edge where the overall agreement between simulation and experiment is not so good. However, the presence of the third peak after the main absorption points towards the existence of a finite amount of Zn$_{Oh}$ in all samples. This provides additional evidence for previous speculations about the presence of Zn$_{Oh}$ which were invoked as source of $p$-type conductivity \cite{PPZ11}. So far, the spectroscopic signature of Zn$_{Oh}$ cannot be used to specify the exact amount, but tentatively the comparison between the central peak height with the two satellite peaks gives a qualitative measure of the amount of Zn$_{Oh}$: the lower the central peak the more octahedral Zn should be present in the sample. Thus, one has to conclude that the presence of Ni facilitates the formation of Zn$_{Oh}$ despite the fact that Ni itself is only incorporated on octahedral sites. To put this spectroscopic measure for the presence of Zn$_{Oh}$ on a more quantitative basis, a well-defined sample series would be required where the amount of Zn on octahedral sites must be known from independent experiments and/or more sophisticated simulations of the spectra are needed. Both however go beyond the scope of this paper. In the following, we turn to the magnetic properties and thereby rely on the following local structure and valencies: Zn is present as 2+ both on octahedral as well as tetrahedral sites, Co$^{3+}_{Oh}$ is dominating while small amounts of tetrahedral Co is evidenced. In contrast Ni is almost exclusively present as Ni$^{3+}_{Oh}$ but with increasing Ni content marked changes are seen which are tentatively assigned to the onset of phase separation of NiO. 

\begin{figure}[tb]
\resizebox{1\columnwidth}{!}{\includegraphics{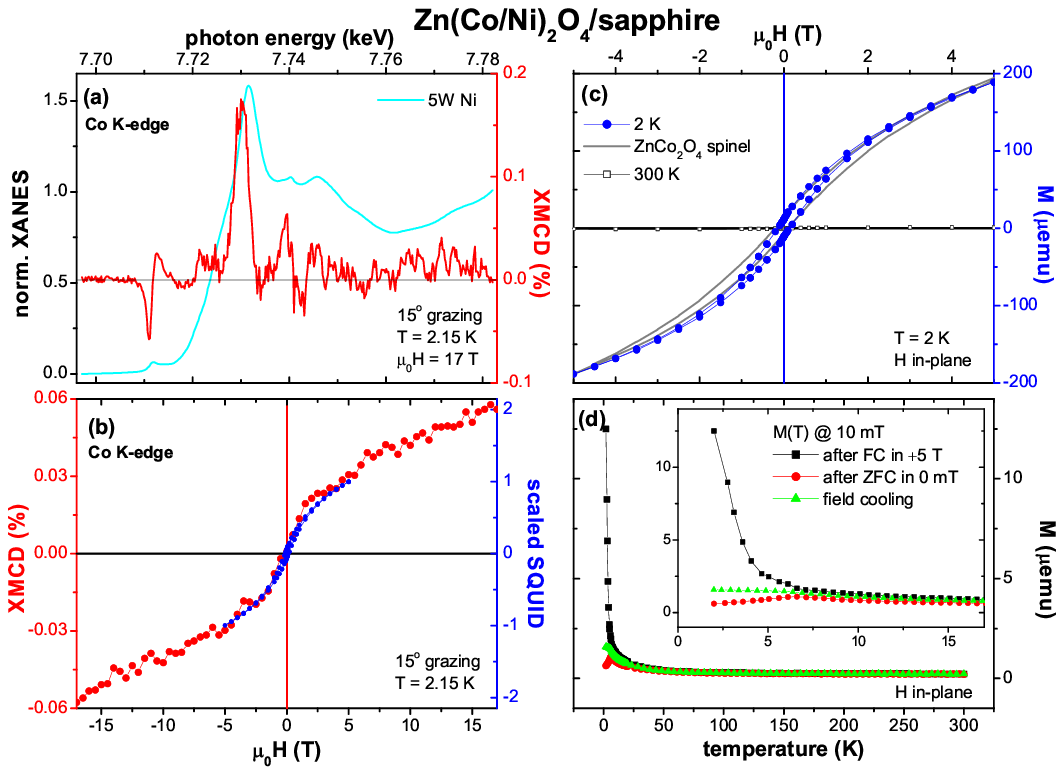}}
\caption{ZnCo$_2$O$_4$ spinel doped with 5~W Ni. (a) shows the resulting XANES and XMCD spectra at the Co $K$-edge; the XMCD$(H)$ behavior is compared to the scaled SQUID $M(H)$ curve in (b). The $M(H)$ curves measured by SQUID is compared to the undoped spinel in (c), the resulting $M(T)$ behavior is shown in (d) while the inset enlarges the low-temperature region. \label{fig10}}
\end{figure}

In Fig.\ \ref{fig10} the magnetic properties of the 5~W Ni sample are compared to the undoped ZnCo$_2$O$_4$ spinel. The Co $K$-edge XMCD in Fig.\ \ref{fig10}(a) is virtually identical to the one of the undoped spinel \cite{HNO15} with a maximum XMCD at the pre-edge of 0.05\% and a characteristic spectral shape. Unfortunately, the amount of Ni was too low so that no reliable Ni $K$-edge XMCD could be recorded. The integral $M(H)$ curves measured by SQUID at 2~K for both samples are shown in Fig.\ \ref{fig10}(b); no magnetic signal can be sensed at 300~K (open squares). Both samples are rather similar and exhibit a narrow opening of a hysteresis over virtually the entire field range of 5~T, no matter if doped with Ni or not; however, a vertical shift of the loop is absent. The only discrepancy between the two samples is that the curvature of the $M(H)$ curve is a bit more rounded for the Ni-codoped sample which may hint towards a slightly larger magnetic moment. The slight opening of a hysteresis can be interpreted as uncompensated antiferromagnetism as done before for Co:ZnO \cite{NHL16}. The difference between the doped wurtzite Co:ZnO and the ZnCo$_2$O$_4$ spinel is, that the mechanism behind the uncompensation has to be different. In the Co:ZnO the uncompensated moments stem from nonmagnetic Zn cation neighbors. In the ideal spinel however, the Co$_{Oh}$ would be fully compensated, so that only the presence of A/B disorder can lead to uncompensated moments, i.\,e., the presence of both Zn$_{Oh}$ as well as Co$_{Td}$ inferred from the XANES analysis in Fig.\ \ref{fig9} can be corroborated by the magnetic properties, in particular the shallow opening of a hysteresis over a wide field range. In Fig.\ \ref{fig10}(c) the XMCD$(H)$-curve at the Co $K$-edge is compared to the integral magnetic properties. The overall shape is rather similar but the high field behavior demonstrates the absence of magnetic saturation up to 17~T, a fact that can be taken as evidence for the presence of antiferromagnetic coupling \cite{NNW12} further substantiating the assumption of uncompensated antiferromagnetic order in the Ni-codoped (as well as the undoped) ZnCo$_2$O$_4$ spinel. Finally, Fig.\ \ref{fig10}(d) summarizes the low-temperature regime of $M(T)$ curves recorded under various cooling conditions. A deviation between the curves is only visible below 7~K which is slightly smaller than for 60\% Co:ZnO and clearly below Co$_3$O$_4$ with its  N\'eel-temperature of 40\,K \cite{Rot64}. 

\begin{figure}[tb]
\resizebox{1\columnwidth}{!}{\includegraphics{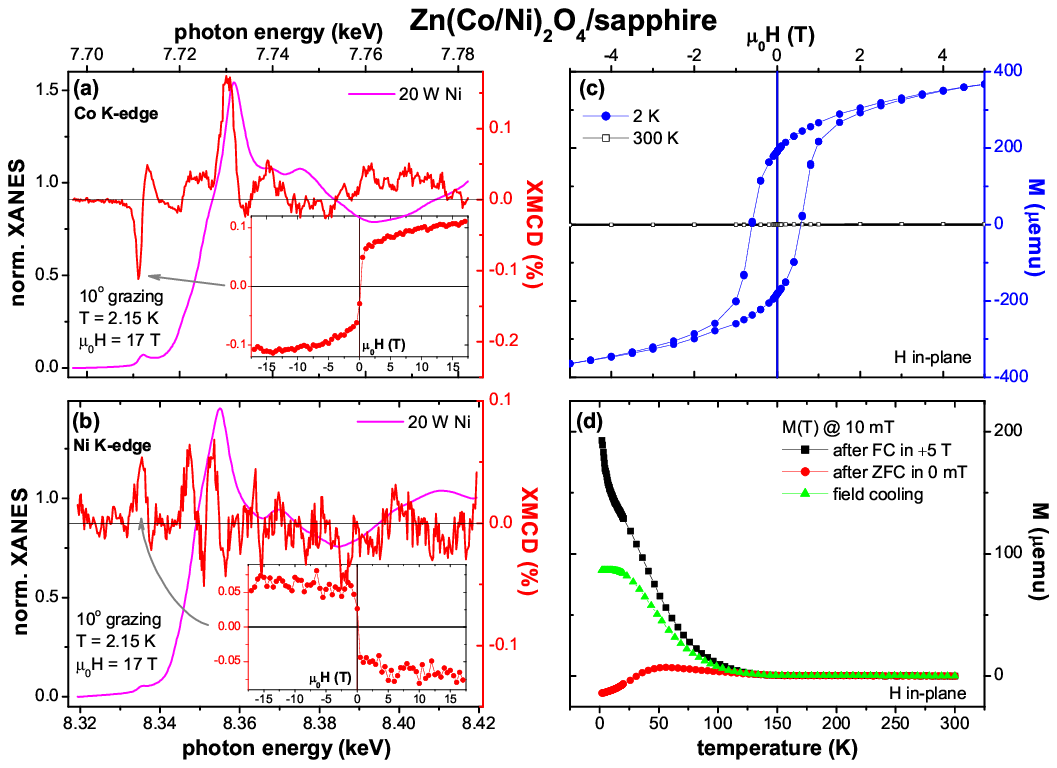}}
\caption{ZnCo$_2$O$_4$ spinel doped with 20~W Ni. XANES and XMCD spectra at the (a) Co and the (b) Ni $K$-edges; the inset shows the respective XMCD$(H)$ behavior measured at the maximum XMCD signal. (c) $M(H)$ curves measured at 2~K by SQUID, the resulting $M(T)$ behavior is shown in (d). \label{fig11}}
\end{figure}

The magnetic properties of the 20~W Ni sample are shown in Fig.\ \ref{fig11}. For this sample, the XMCD could be recorded at the Co (a) as well as the Ni (b) $K$-edges and the respective insets shown the respective XMCD$(H)$ behavior. On the full field scale of 17~T no opening of a hysteresis is visible but around zero field a pronounced jump of the XMCD signal can be seen at either edge. More remarkable, the sign of the XMCD signal at the Ni $K$-edge is reversed compared to the Co $K$-edge, which demonstrates that the magnetic moments are aligned antiparallel. This can also explain why the XMCD signal at the Co $K$-edge is doubled to 0.1\% compared to the 5~W sample. Introducing Ni with its smaller atomic moment in an antiferromagnetically coupled Co sublattice with its larger atomic moment leads to a further uncompensation of the moments and thus an overall increase in the effective magnetic moment. In that regard also the element specific magnetic properties corroborate the findings of the structural characterization in Fig.\ \ref{fig9}; also the findings are similar to those in Co/Ni codoped ZnO in Fig.\ \ref{fig7a} the incorporation of Ni increases the effective magnetic moment measured at the Co edge but here we have direct experimental evidence for the antiferromagnetic coupling. Note that this interpretation requires only considering the octahedral sublattice of the spinel; however, the presence of the shallow hysteresis in the undoped spinel, which is also seen in the 5~W Ni codoped sample, requires the presence of Co$_{Td}$ in addition, which according to the Co $K$-edge XANES is also present in the 20~W Ni sample. 

The integral magnetic properties of the 20~W Ni sample were recorded using SQUID and the $M(H)$-curve at 2~K in shown in Fig.\ \ref{fig10}(c) while Fig.\ \ref{fig10}(d) summarizes the $M(T)$-curves under various cooling conditions. In stark contrast to the 5~W sample, a clear hysteresis is visible and the magnetic order extends to above 100~K. This behavior, together with the results of the XMCD are indicative of ferrimagnetism rather than uncompensated antiferromagnetism; however, both variants are closely interrelated. While the more pronounced hysteresis can be made plausible by the larger effective magnetic moment the increased temperature regime is more difficult to be understood. For NiCo$_2$O$_4$ which is an inverse spinel, ferrimagnetism with a high Curie temperature of over 600~K is known and recent work reports a coercive field of about 0.7~T at 10~K \cite{BCL15}. The latter is of the same order as in the present sample, however, the magnetic order is restricted to much lower temperatures. Obviously, the presence of Zn significantly weakens the magnetic interactions compared to the NiCo$_2$O$_4$ spinel. At present we cannot conclude on the microscopic origin of the magnetic interactions in the Ni codoped ZnCo$_2$O$_4$ spinel, but the fact that the temperature regime of magnetic order is reduced that strongly may be associated with the fact, that Zn is incorporated on both tetrahedral as well as octahedral sites. Thereby, it can influence all relevant sublattice exchange couplings, i.\,e., the Oh-Oh, the Td-Td and the Td-Oh coupling. If we take the latter to be negligible in the present case (only little Co$_{Td}$ and none Ni), the Td-Td exchange should also play no significant role. However, the Oh-Oh interaction is antiferromagnetic and weak, as seen in the undoped ZnCo$_2$O$_4$ spinel as well as the 5~W Ni sample. Thus, the presence of a significant amount of Co$_{Td}$ is required to account for the high Curie temperature of NiCo$_2$O$_4$. In turn, the low occurrence of Co$_{Td}$ seems to be key to understand the lower order temperature of the 20~W Ni sample compared to NiCo$_2$O$_4$. In turn, for the increase of the order temperature compared to the 5~W the significantly increased Ni$_{Oh}$ content appears to strengthen the magnetic properties; in other words, the magnetic Oh-Oh exchange seems to be more robust for Ni compared to Co. However, since the Ni $K$-edge XANES is altered and a small shoulder in the XRD appears, it is difficult to draw more explicit conclusions beyond the above qualitative arguments.  

\begin{figure}[tb]
\resizebox{1\columnwidth}{!}{\includegraphics{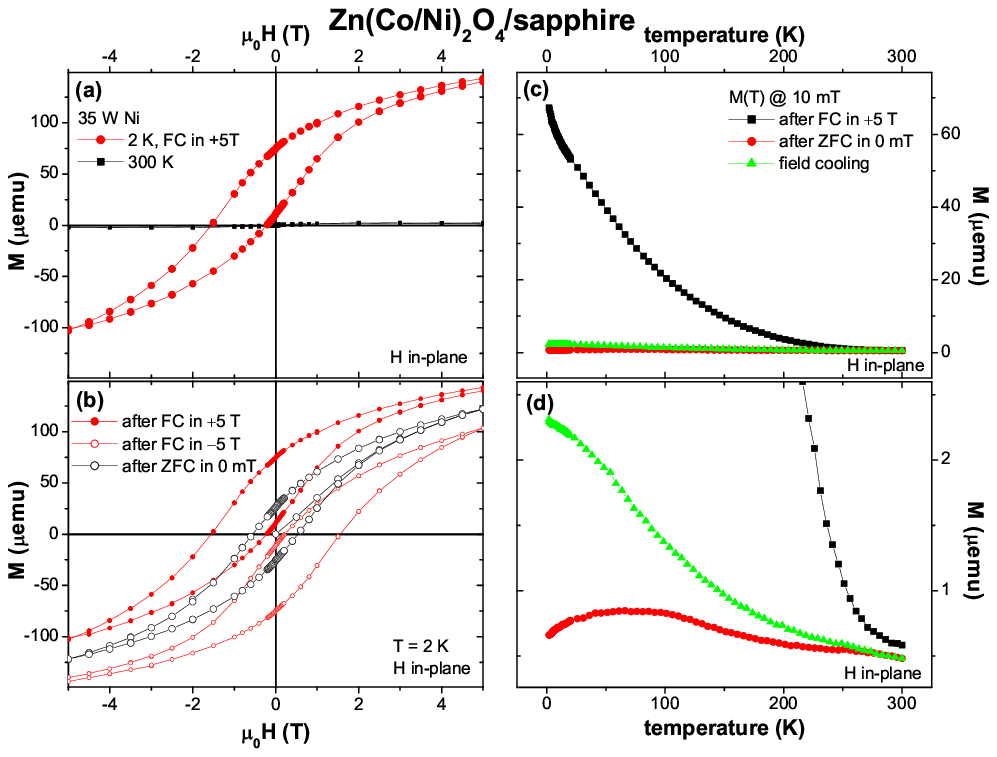}}
\caption{ZnCo$_2$O$_4$ spinel doped with 35~W Ni. (a) displays the $M(H)$ behavior at 2~K and 300~K measured by SQUID, while (b) summarizes the cooling-field dependence of the $M(H)$ curve at 2~K. The resulting $M(T)$ curves are shown in (c) and (d), respectively.  \label{fig12}}
\end{figure}

Finally, the 35~W Ni sample shall be discussed with regard to its magnetic properties. Figure \ref{fig12} summarized the integral magnetic properties as measured with SQUID: The $M(H)$-curve at 2~K shown in Fig.\ \ref{fig12}(a) was recorded after cooling in a field of 5~T and a clear exchange bias is visible. The cooling field dependence of the 2~K $M(H)$ loop is depicted in Fig.\ \ref{fig12}(b) and in comparison to cooling in zero field (ZFC, open symbols) the $\pm 5$~T cooling field induced both a vertical as well as a horizontal shift of the loop. Since for this sample the presence of NiO has been evidenced by XRD, this behavior is indicative of an exchange bias system where the most likely antiferromagnet is NiO while the ferro-/ferrimagnetic counterpart is the above-discussed Ni-codoped ZnCo$_2$O$_4$ spinel. The latter assignment is based only on the coercivity of the ZFC hysteresis which is very comparable to the one of the 20~W Ni sample and the similarity of the Zn and Co $K$-edges seen in Fig.\ \ref{fig8}(b) and (c). Note that the amount of tetrahedral Zn appears to be a bit higher in the 35~K Ni sample because the central peak of the XANES is more pronounced compared to the 5~W and 20~W Ni samples. The changes at the Ni $K$-edge in turn is explained by the formation of the NiO secondary phase. While the actual micro-/mesoscopic configuration of the sample cannot be clarified based on the presented experiments, such an exchange-biased composite oxide material may be interesting by itself. For the purpose of the present paper the magnetic properties are taken as another independent experimental evidence for the formation of a secondary phase. Note, that the $M(T)$ curves in Fig.\ \ref{fig12}(c) show the presence of an antiferromagnetic order up to temperatures of close to 300~K. This is almost identical with the behavior of the 10\% Ni:10\% Co:ZnO sample shown in Fig.\ref{fig7}(b), thus the potential secondary phase appears to be very comparable for both rather different types of sample. The different here is, that the remaining part of the sample is not an anisotropic paramagnet but a ferrimagnet similar to the one shown in Fig.\ \ref{fig11}. Note, that at the order temperature of the 20~W Ni sample around 100~K there is a small kink in the $M(T)$-curves upon field cooling or after ZFC, which are enlarged in Fig.\ \ref{fig12}(d). The remaining, more narrow separation between the two curves between 100~K and close to 300~K may thus be induced/stabilized by the presence of the antiferromagnetic component, which loses its magnetic order just around 300~K.

\section{Conclusion}

In conclusion, we have shown for a comprehensive set of samples that the solubility of Ni in wurtzite ZnO, Co:ZnO as well as in the ZnCo$_2$O$_2$ spinel is rather limited. In the wurtzite environment almost exclusively Ni$^{2+}_{Td}$ is formed and Ni$^{3+}$ can only be created by photo-ionization. An increased conductivity upon Ni diffusion can only be explained by indirect effects of the Ni incorporation via other, unintentional defects. Eventually, phase separation of presumably NiO sets in when a Ni concentration of about 10\% is reached. The substitutional Ni in wurtzite ZnO exhibits an opposite single-ion anisotropy compared to Co and in the codoped sample evidence for an antiparallel magnetic coupling of Ni and Co moments is found. In the spinel, predominantly Ni$^{3+}_{Oh}$ is formed. The Ni moments couple antiparallel to those of Co which itself mostly exist as Co$^{3+}_{Oh}$ but a small amount of tetrahedral Co is seen. This gives rise to an uncompensated antiferromagnetism which is further uncompensated by the addition of Ni when the sample turns ferrimagnetic with a magnetic order which extends to higher temperatures, around 100~K. Further increasing the Ni content again leads to phase separation of NiO rendering the sample to exhibit exchange bias. Finally, in all spinel samples there are indications of the previously invoked presence of Zn$^{2+}_{Oh}$ and a characteristic spectroscopic signature in the Zn $K$-edge XANES has been suggested based upon simulations and comparison to reference spectra. Especially in the spinel samples such types of reference spectra will be very useful to understand interesting physical and/or chemical properties on a microscopic level.  

\section*{Acknowledgment}

The authors gratefully acknowledge funding by the Austrian Science Fund (FWF) - Project No. P26164-N20. K. M. Johansen would like acknowledge the Research Council of Norway for the support to the Norwegian Micro- and Nano-Fabrication Facility, NorFab, project number 295864 and GO-POW, project number 314017.

\end{document}